\documentclass[journal]{IEEEtran}
\usepackage{cite}

\ifCLASSINFOpdf
   \usepackage[pdftex]{graphicx}
  \graphicspath{{figures/}}
   \DeclareGraphicsExtensions{.pdf,.jpeg,.png}
\else
  \usepackage[dvips]{graphicx}
  \graphicspath{{figures/}}
  \DeclareGraphicsExtensions{.eps,.jpeg}
\fi
\usepackage[cmex10]{amsmath}
\ifCLASSOPTIONcompsoc
  \usepackage[caption=false,font=normalsize,labelfont=sf,textfont=sf]{subfig}
\else
  \usepackage[caption=false,font=footnotesize]{subfig}
\fi
\usepackage{url}

\usepackage{amsfonts}
\usepackage{graphicx}
\usepackage{mathtools}
\usepackage{enumerate}
\usepackage{multirow}
\usepackage{todonotes}
\usepackage{array}
\usepackage{tabu}
\usepackage{booktabs}
\usepackage[caption=false,font=footnotesize]{subfig}
\usepackage{bm}
\usepackage{caption}

\begin{document}

\title{Sound Context Classification Basing on Join Learning Model and Multi-Spectrogram Features}
%
\author{Dat~Ngo,
	Hao~Hoang,
	Anh~Nguyen, 
	Tien~Ly, 
    Lam~Pham
\thanks{Dat~Ngo, Hao~Hoang, Anh~Nguyen are with the Electrical and Electronics Department, Ho Chi Minh City University of Technology, Vietnam.}
\thanks{Tien~Ly is with University of Nottingham, UK.}
\thanks{Lam~Pham is with the School of Computing, University of Kent, UK.}%
\thanks{All authors have same contribution on this paper.}%

%
}

\markboth{XX,~Vol.~xx, No.~X, May~20XX}%
{Dat Ngo\MakeLowercase{\textit{et al.}}: IEEE Transactions on Audio, Speech and Language Processing}

\maketitle
\begin{abstract}
In this paper, we present a deep learning framework applied for Acoustic Scene Classification (ASC), the task of classifying scene contexts from environmental input sounds. 
An ASC system generally comprises of two main steps, referred to as front-end feature extraction and back-end classification.
In the first step, an extractor is used to extract low-level features from raw audio signals. 
Next, the discriminative features extracted are fed into and classified by a classifier,reporting accuracy results.
Aim to develop a robust framework applied for ASC, we address exited issues of both the front-end and back-end components in an ASC system, thus present three main contributions: 		
Firstly, we carry out a comprehensive analysis of spectrogram representation extracted from sound scene input, thus propose the best multi-spectrogram combinations. 
In terms of back-end classification, we propose a novel join learning architecture using parallel convolutional recurrent networks, which is effective to learn spatial features and temporal sequences of spectrogram input.
Finally, good experimental results obtained over benchmark datasets of IEEE AASP Challenge on Detection and Classification of Acoustic Scenes and Events (DCASE) 2016 Task 1, 2017 Task 1, 2018 Task 1A \& 1B, LITIS Rouen prove our proposed framework general and robust for ASC task.
\end{abstract}
%
\section{Introduction}
\label{sec:introduction}

Acoustic Scene Classification (ASC), which aims to identify a sound scene context, vitally contributes to a variety of real-life applications ranging from security \cite{zwan2008automatic}, surveillance \cite{valero2012gammatone} and context-aware consumer services \cite{schilit1994context, heittola2013context, xu2008intelligent}.	
Although ASC research is very close to Automatic Speech Recognition (ASR) and Speaker Recognition System (SRS) due to exploring audio signals, ASC currently presents various and different challenges.
Firstly, there are a wide range of acoustic events in real-world environments, and these occur in different ways.  
Some sound events constitute natural auditory scenes that presents an acoustic mixture signal.
For instance, \textit{bird} sounds and the sounds of \textit{leaves, grass, or trees blowing} in the wind clearly indicate certain context like \textit{in a park} or \textit{on a field}. 
However, it is more difficult to handle some sound events that are not context specific such as  \textit{engine}, or \textit{talking}. 
This kind of context causes a confusion for even human (only listening) to recognize exactly \textit{in a street, on a transportation such as car, bus or tube} or \textit{in a station}. 
Indeed, experimental results in ~\cite{ma2003context} indicates that a proposed ASC system configured by Mel Frequency Cepstral Coefficients (MFCCs)-Hidden Markov Model (HMM) significantly outperformed human ability for recognizing everyday acoustic.
Secondly, if sound events are considered as signal that mixed in diverse scenes as noise, there are different levels of signal-to-noise ratio (SNR) due to environmental conditions, distance of recording devices and so on. 
Moreover, these sounds exist across a wide range of frequency bands. 
Some occupy narrow frequency bands, while some spread over wide bands, and many sounds have frequency bands that overlap each other. 
Finally, natural sounds in ASC research do not follow any structure, unlike a speech signal. 

To deal with these challenges, the state-of-the-art systems tend to make use of multi-input features.
In particular, systems approaching frame-based features make effort to combine frequency and temporal features to maximize the chance of correct feature representation. 
For instance, MFCCs \cite{mcloughlin2016speech}, one of most used frquency features, is combined with a wide range of temporal features such as loudness, average short-time energy, sub-band energy, zero-crossing rate, spectral flux, or spectral centroid in \cite{marchi2016pairwise, marchi2016up, vafeiadis2017acoustic}. 
Similarly, an effective combination of MFCCs with a variety of features such as perceptual linear prediction (PLP) coefficients, power nomalised cepstral coefficients (PNCC), robust compressive gamma-chirp filter-bank cepstral coefficients (RCGCC) or subspace projection cepstral coefficients (SPPCC) was proposed in \cite{park2016score} that helps to achieve the top-three system in DCASE 2016 challenge. 
As usual, ASC systems approaching frame-based feature representation use traditional machine learning models for back-end classification, such as Hidden Markov Model (HMM) \cite{ma2003context}, Support Vector Machine (SVM) \cite{geiger2011learning, geiger2013large}, and  Gaussian Mixture Model (GMM) \cite{vuegen2013mfcc}.

Inspire that frame-based representation may not capture enough information, two-dimensional spectrograms as a more effective way for low-level feature representation have been exploited by the state-of-the-art ASC systems. 	
In particular, spectrograms such as short term Fourier transform (STFT) \cite{mcloughlin2016speech}, log-Mel \cite{ren2019attention, phaye2019subspectralnet}, MFCC \cite{mesaros2017dcase}, constant-Q transform (CQT) \cite{lidy2016cqt}, and Gammatone spectrograms (GAM) \cite{phan2017improved, phan2017audio} are the most frequent low-level feature used.
To further enrich input features, multiple spectrograms are widely approached. 
For instances, log-Mel is combined with a different types of spectrograms such as Mel-based nearest neighbour filter (NNF) spectrogram \cite{nguyen2018acoustic, nguyen2019acoustic}, CQT \cite{zeinali2018convolutional}, or two spectrograms such as MFCC and GAM in \cite{phan2017improved, phan2016label}. 
Furthermore, the idea of generating multi-spectrogram input is also based on combinations of information from channels of recording devices. 
This idea has been proved its efficiency when Yuma \textit{et al.} \cite{sakashita2018acoustic} did experiments to explore the average and difference of two channels, separated harmonic and percussive spectrograms from each channel, consequently achieved the top-one score in DCASE 2018 Task 1A challenge.
Approach both frame-based and spectrogram representation, combined features such as MFCC+i-Vector or log-Mel+x-Vector were proposed in \cite{jung2018dnn} and \cite{zeinali2018convolutional}, respectively.	
To explore two-dimensional spectrogram representation, ASC systems usually deploy complicated classification models, mainly coming from deep learning techniques.
For examples, Yang \textit{et al.} \cite{yang2018acoustic} proposed a complicated CNN-based architecture called the \textit{Xception} network. 
This is inspired by the fact that a deep learning network trained by a wide range of feature scales and over separated channels can result in a very powerful model. 
Besides, Truc \textit{et al.} \cite{nguyen2018acoustic} applied a parallel CNNs to learn from two types of spectrogram (log-Mel and NNF).
Next, the two outputs of the CNNs are concatenated to generate high-performed features that were thus explored by an DNN and achieve the highest accuracy rate in DCASE 2018 Task 1B challenge. 
Approach Recurrent Neural Network (RNN) based architecture, Zang \textit{et al.} \cite{zhang2018temporal,zhang2018data, zhang2018multi} provided a deep analysis of the application of Long Short-Term Memory (LSTM) for ASC. 
Other examples prove effective in exploiting RNN-based networks for ASC were published by Huy \textit{et al.} \cite{phan2017audio, phan2019spatio, phan2018beyond}. 
Instead of using LSTM, Huy \textit{et al.} proposed using Gate Recurrent Unit (GRU-based) based architecture \cite{phan2019spatio}. 
Then they further improved the model by applying an attention scheme \cite{phan2017audio} or combining with a CNN-based architecture \cite{phan2018beyond}.
Compare to frame-based approach, ASC systems using spectrogram representation outperform and show more robust \cite{phan2017improved, phan2019spatio, phan2018beyond}.

As the analysis of the-state-of-the-art ASC systems, we adopt the second trend that uses spectrogram representation for low-level feature input and explore deep learning model architecture for classification.
In particular, we mainly contribute:
\begin{itemize}
	\item Although spectrogram-based ASC systems explore multi-spectrogram input features to deal with ASC challenges, none of research has analysed and indicated the most effective combination of spectrograms.
	In this paper, we, therefore, provide a comprehensive analysis on spectrograms by conducting experiments on five common types of spectrograms, comprising of Short-time Fourier Transform (STFT), log-Mel, Mel Frequency Ceptral Coefficient (MFCC), Constant Q Transform (CQT), and Gammatone filter (GAM).
	To do this, we firstly introduce a C-DNN-based deep learning model, likely VGG-9 \cite{gousseau2019vgg}.
	Consequently, we evaluate individual spectrograms on C-DNN network proposed, thus indicate the most effective combination of spectrograms due to late fusion of individual spectrogram accuracy.

	\item Next, we improve the C-DNN model by adding a parallel C-RNN based architecture to efficiently learn the structure of temporal sequences of spectrograms. 
	By using a parallel CNN and C-RNN networks, we create a join learning architecture that is very useful to deploy the two-dimensional spectrogram input.		
	
	\item To evaluate ASC systems, researches normally did experiments on one dataset \cite{marchi2016up, park2016score} . 
		Some proposed to evaluate on two datasets \cite{li2018acoustic,ren2019attention}. 
		This may not conclude ASC systems proposed general or powerful. 
		We, therefore, conduct extensive experiments, evaluating our proposed systems over five ASC datasets of DCASE 2016 Task 1, DCASE 2017 Task 1, DCASE 2018 Task 1A \& 1B,  and Litis Rouen published recently.
		Good results obtained from experiments on various datasets showing different category number, recording time, and wide range of real-life environments strongly prove the proposed system general and robust.
\end{itemize}

%

\section{Dataset and Setting}
\label{sec:Data_Set}

Our experiments are conducted over a variety of published ASC datasets, comprising of LITIS Rouen~\cite{rakotomamonjy2014histogram}  and IEEE AASP Challenge on Detection and Classification of Acoustic Scenes and Events (DCASE) 2016 Task 1~\cite{Mesaros2016_EUSIPCO} , 2017 Task 1 ~\cite{mesaros2017dcase} , 2018 Task 1A \& 1B \cite{Mesaros2018_DCASE}.

\textbf{LITIS Rouen} dataset was recorded at a sample rate of 22050 Hz with 3026 segments, each presents 30-s duration.
This dataset contains totally 25.51 recording hours for 19 urban scene categories, showing unbalanced data. 
Following the mandated settings, the dataset is separated and organised for 20-fold cross validation, reporting the final classification accuracy by averaging over the 20 testing folds.

\textbf{DCASE} 2016 Task 1 and DCASE 2017 Task 1 similarly present 15 categories and were recorded at 44.1 kHz.
	While each segment in DCASE 2016 is  30\,s, 10-s duration is presented in DCASE 2017.
	Noticeably, DCASE 2017 reuses all DCASE 2016 and adds new data recorded. 
	Obey the recommended setting, we train our proposed system on development set (Dev.) and evaluate on the evaluation set (Eva.).
	As regards DCASE 2018 Task 1A, it was recorded at 44.8 kHz, spanning 10 categories and used one recording device namely A.
	DCASE 2018 Task 1B reuses all data from DCASE 2018 Task 1A, and adds more data recorded by two different devices, namely B and C.
	Noticeably, the total recording time conducted on device B and C are much less than device A (denoted as DCASE 2018 Task 1A), reporting totally 4 hours  on B\&C compared to 24 hours in device A. 
	As a result, DCASE 2018 Task 1B dataset involves issues of mismatched recording devices and unbalanced data in terms of recording devices. 
	Therefore, DCASE 2018 Task 1B challenge only compare systems' results on device B\&C with less recording time.
	As DCASE 2018 Task 1A and 1B have not released labels of evaluation set, we separate development set into two sub-sets, namely Training and Test sets for training and testing processes respectively.
	While DCASE 2016 Task 1 and DCASE 2017 Task 1 are balanced, little unbalanced data is shown in DCASE 2018 Task 1A and 1B.

\section{High-level architecture}
\label{sec:high_level_arc}
\begin{figure*}[ht]
	\centering
	\centerline{\includegraphics[width=\linewidth]{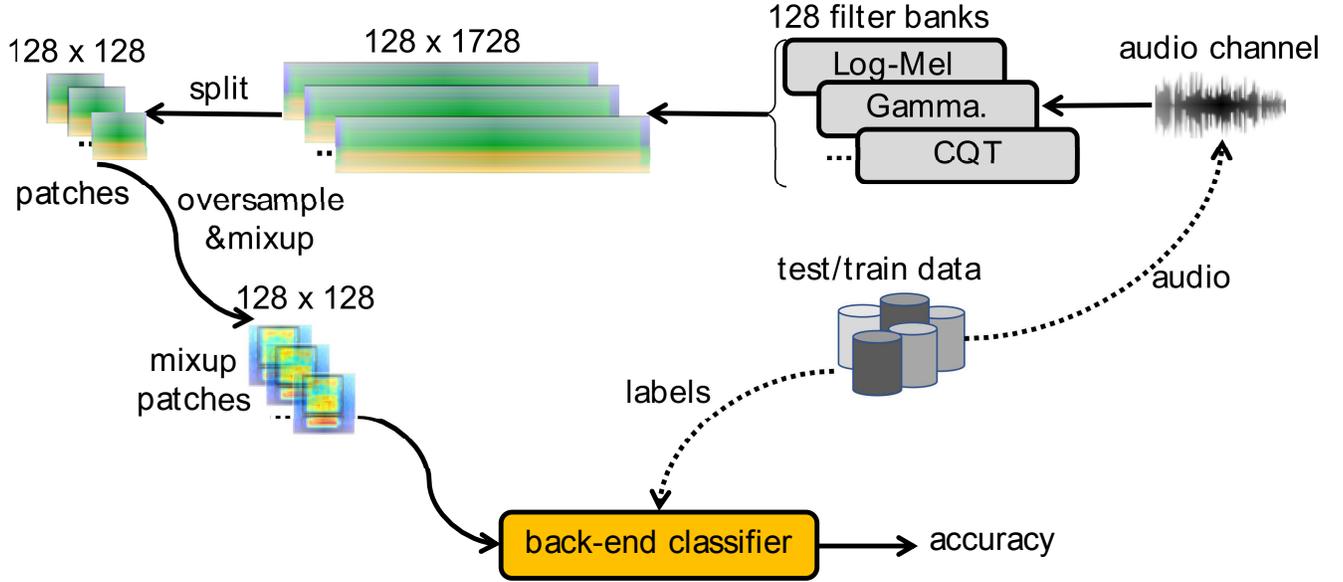}}
	\vspace{0.1cm}
	\caption{High-level architecture of our ASC system.}
	\label{fig:high_level_arc}
\end{figure*}

Our proposed deep learning framework applied for ASC, in general, is described in Figure \ref{fig:high_level_arc}.
As Figure \ref{fig:high_level_arc} shown, the framework is separated into low-level feature extraction (the upper part) and back-end classification (the lower part). 
In particular, the draw audio from the Channel 1 is firstly transformed into spectrogram representation, using 128 filter banks.
The entire spectrogram is thus split into non-overlapped image patches of $128\times128$.
To deal with unbalanced data issue, we apply two data augmentation techniques on the image patches.
	Firstly, we randomly oversample image patches which belong in categories with less audio segments.
	Next, the mixup data augmentation \cite{salamon2017deep} is applied, thus generate new image patches.
Let's consider two original image patches as \(\mathbf{X_{1}}\), \(\mathbf{X_{2}}\) and expected labels as \(\mathbf{y_{1}}\), \(\mathbf{y_{2}}\), new image patches are generated as below equations:
\begin{equation}
\label{eq:mix_up_x1}
\mathbf{X_{mp1}} = \mathbf{X_{1}}*\gamma + \mathbf{X_{2}}*(1-\gamma) 
\end{equation}
\begin{equation}
\label{eq:mix_up_x2}
\mathbf{X_{mp2}} = \mathbf{X_{1}}*(1-\gamma) + \mathbf{X_{2}}*\gamma
\end{equation}
\begin{equation}
\label{eq:mix_up_y}
\mathbf{y_{mp1}} = \mathbf{y_{1}}*\gamma + \mathbf{y_{2}}*(1-\gamma)
\end{equation}
\begin{equation}
\label{eq:mix_up_y}
\mathbf{y_{mp2}} = \mathbf{y_{1}}*(1-\gamma) + \mathbf{y_{2}}*\gamma
\end{equation}
where \(\gamma\) is random coefficient from both \textit{unit} and \textit{beta} distribution, \(\mathbf{X_{mp1}}\), \(\mathbf{X_{mp2}}\) and \(\mathbf{y_{mp1}}\), \(\mathbf{y_{mp2}}\) are new image patches and labels generated, respectively.
Eventually, the mixup patches are fed into a back-end classifier, report the classification accuracy.

\section{An analysis of spectrogram features}
\label{sec:spectrogram_analysis}

By using the high-level architecture mentioned above, we evaluate five individual spectrograms (STFT, log-Mel, MFCC, GAM, and CQT), thus indicate which kind of spectrograms and their combinations is the most influencing on our system's performance.
To evaluate the individual spectrograms, we firstly proposed a C-DNN based network referred to as the back-end classifier. 
Next, we create five systems, each reuse the high-level framework architecture mentioned with one type of spectrogram and C-DNN architecture for classification.
Noting that we use same setting with window size=1290, hop size=256, frequency minimum\(f_{min}\)=10 Hz, and filter bank number=128 to generate same-size spectrograms.
The results of five systems over DCASE 2018 Task 1B dataset are analysed and compare to the DCASE baseline.
Inspire that each spectrogram contain discriminative and complementary features, we fuse the individual systems' accuracy results, thus indicate which combination of spectrograms is effective to improve the performance.

\subsection{C-DNN architecture proposed for back-end classification}
\label{ssec:Baseline Architecture}
\begin{figure*}[h]
	\centering
	\captionsetup{justification=centering}
	\centerline{\includegraphics[width=\linewidth]{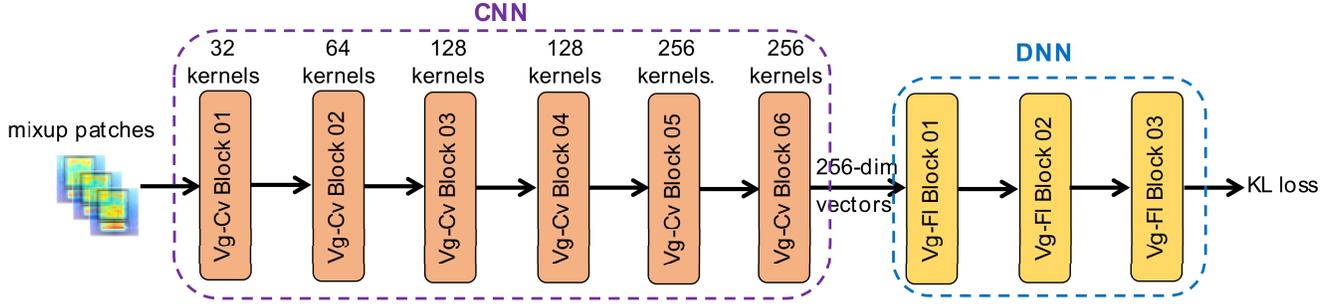}}
	\caption{Block-level architecture of C-DNN network.}
	\label{fig:baseline_architecture}
\end{figure*}
\begin{table*}[h]
	\caption{Network layers used in C-DNN architecture.} 
	\vspace{0.01cm}
	\centering
	\begin{tabular}{c l l c} 
		\toprule 
		& \textbf{Blocks}   &  \textbf{Layers}  &   \textbf{Outptut shape}  \\
		\midrule 
		&  Input  &                                                                                                  &  $128{\times}128{\times}1$ \\       
		&Vg-Cv Block 01 &  Bn - Cv [$9\times9$] - ReLu - Bn - Ap [$2\times2$] - Dr (10\%)  & $64{\times}64{\times}32$        \\
		&Vg-Cv Block 02 &  Cv [$7\times7$] - ReLu - Bn - Ap [$2\times2$] - Dr (15\%)  & $32{\times}32{\times}64$        \\
		\textbf{CNN} &Vg-Cv Block 03 &  Cv [$5\times5$] - ReLu - Bn - Dr (20\%)  & $32{\times}32{\times}128$        \\
		\textbf{Architecture}&Vg-Cv Block 04 &  Cv [$5\times5$] - ReLu - Bn - Ap [$2\times2$] - Dr (20\%)  & $16{\times}16{\times}128$        \\
		&Vg-Cv Block 05 &  Cv [$3\times3$] - ReLu - Bn - Dr (25\%)  & $16{\times}16{\times}256$        \\
		&Vg-Cv Block 06 &  Cv [$3\times3$] - ReLu - Bn - Gap - Dr (25\%)  & $256$        \\
		\bottomrule     
		&Vg-Fl Block 01 &  Fl - ReLu - Dr (30\%)  & $512$        \\
		\textbf{DNN} &Vg-Fl Block 02 &  Fl - ReLu - Dr (30\%)  & $1024$        \\
		\textbf{Architecture}&Vg-Fl Block 03 &  Fl - Softmax & 10        \\
		\bottomrule 
	\end{tabular} 
	\label{table:cnn_arc} 
\end{table*}

In order to evaluate individual and multiple spectrograms, we proposed a C-DNN network architecture as described in Figure \ref{fig:baseline_architecture} and Table \ref{table:cnn_arc}.
As Figure \ref{fig:baseline_architecture} shown, C-DNN architecture comprises of CNN and DNN parts in order.
CNN part is described by 6 Vg-Cv blocks, performed by  Batchnorm (Bn), Convolutional (Cv[kernel size]),  Rectified linear unit (ReLu), Dropout (Dr(Percentage dropped)), Average Pooling (Ap), Global Average Pooling (Gap) layers as showed in the top of Table \ref{table:cnn_arc}.
Meanwhile, DNN part in Figure \ref{fig:baseline_architecture} is configured by three Vg-Fl blocks with Fully-connect (Fl),    ReLu, Dropout (Dr(Percentage dropped)), and Softmax layers, as described in the bottom of Table \ref{table:cnn_arc}.
It can be seen that CNN part helps to map input image patches to condensed and discriminative vectors, referred to as high-level features.
Each high-level feature vector presents 256 dimensions due to the number of kernels used in the final convolutional layer in Vg-Cv block 06.
Next, DNN part explores the high-level features, thus classifies into 10 categories (the category number in DCASE 2018 Task 1B dataset evaluated) and reports the classification accuracy.

\subsection{Spectrogram representation and their combinations proposed}
\label{ssec:spectrogram representation}

To propose spectrogram combinations, formulas of individual spectrograms are firstly presented below: 
 
\textbf{Short-Time Fourier Transform (STFT):} The first STFT spectrogram evaluated applies Fourier Transform to extract Frequency content of local section of input signal over short time duration.
Let consider $\mathbf{s}$($n$) as digital audio signal with length of \(N\) , a pixel value at central frequency \(f\) and time frame \(t\) of STFT spectrogram \(\mathbf{STFT}[F,T]\) is computed as:
\begin{equation}
\label{eq:stft}
\mathbf{STFT}[f,t] = \sum_{n=0}^{N-1} \mathbf{s}[n].\mathbf{w}[t]e^{-j2{\pi}fn}  
\end{equation}
where $\mathbf{w}$[$t$] is a window function, typically Hamming.
While time resolution ($T$) of STFT spectrogram is set by window side and hope size, the frequency resolution ($F$) equals to the number of central frequencies set to 2048. 
The frequency resolution, eventually, re-scales into 128 that is same as other spectrograms.

\textbf{log-Mel}: To generate log-Mel spectrogram, draw audio signal is firstly transformed into STFT spectrogram recently mentioned.
Next, a Mel filter bank, which simulates the overall frequency selectivity of the human auditory system using the frequency warping $F_{mel} = 2595.log(1 + {F}/{700})$~\cite{mcloughlin2016speech}, is applied to generate a Mel spectrogram $\mathbf{MEL}[F_{mel}, T]$ (noting that frequency resolution ($F_{mel}$) is the Mel filter number set to 128).
Eventually, logarithmic scaling is applied to obtain the log-Mel spectrogram. 
Let consider $\mathbf{COE}[F_{mel}, F]$ as matrix storing coefficients of Mel filters, log-Mel spectrogram likely a matrix is computed by: 


\setlength{\arraycolsep}{0.0em}
\begin{eqnarray}
\label{eq:log-Mel}
\mathbf{log{-}Mel}[F_{mel},T]&{}={}&log\mathbf{COE}[F_{mel}, F]\nonumber\\
&&.\mathbf{STFT}[F, T]
\end{eqnarray}
\setlength{\arraycolsep}{5pt}


\textbf{Mel Frequency Cepstral Coefficient (MFCC):} From log-Mel spectrogram, Discrete Cosine Transform (DCT) is used to extract a sequence of uncorrelated coefficients crossing frequency dimension, reducing log-Mel frequency resolution into smaller space.  A pixel value $\mathbf{DCT}[f_{dct}, t_{dct}]$ of DCT matrix $\mathbf{DCT}[F_{dct}, T_{dct}]$, where $F_{dct}$ and $T_{dct}$ are frequency and time resolutions, is computed by:

\setlength{\arraycolsep}{0.0em}
\begin{eqnarray}
\label{eq:dct}
\mathbf{DCT}[f_{dct}, t_{dct}]&{}={}&\left(   \frac{2}{F_{mel}} \right)^{\frac{1}{2}}. \left( \frac{2}{T} \right)^{\frac{1}{2}} .\sum_{f_{mel}=0}^{F_{mel}-1}.\sum_{t=0}^{T-1}\nonumber\\
&&\Lambda(f_{mel}).cos\left[ \frac{{\pi}f_{dct}}{F_{mel}}(2f_{mel}+ 1) \right]\nonumber \\
&& .\Lambda(t).cos\left[ \frac{{\pi}t_{dct}}{T}(2t+1)   \right]\nonumber \\
&& .\mathbf{log{-}Mel}[f_{mel}, t]
\end{eqnarray}
\setlength{\arraycolsep}{5pt}

where
\begin{equation}
\label{eq:lamda}
\Lambda(x) = 
\begin{cases}
\frac{1}{\sqrt{2}} & \text{if } x = 0  \\
1 & \text{otherwise}
\end{cases}
\end{equation}
$T$ and $F_{mel}$ are time and frequency resolution of log-Mel spectrogram. 

Next, delta coefficients per time frame showing difference of DCT coefficients over time are computed, shown in Equation. (\ref{eq:delta}).

\setlength{\arraycolsep}{0.0em}
\begin{eqnarray}
\label{eq:delta}
\mathbf{DELTA}[F_{dct}, t] &{}={}&\frac{1}{2} (\mathbf{DCT}[F_{dct}, t-1]\nonumber\\
&& - \mathbf{DCT}[F_{dct}, t+1])\
\end{eqnarray}
\setlength{\arraycolsep}{5pt}

Eventually, $\mathbf{DELTA}[F_{dct}, T_{dct}]$ is concatenated with DCT spectrogram $\mathbf{DCT}[F_{dct}, T_{dct}]$ across frequency dimension to generate MFCC spectrogram as expression $\mathbf{MFCC}[F_{mfcc}, T_{dct}]$ (noting that MFCC frequency resolution ($F_{mfcc}$) doubles frequency resolution of DCT ($F_{dct}$) and equals to 128, and $T_{dct}$ is set to equal to $T$ resolution of log-Mel spectrogram).

\textbf{Constant Q transform (CQT):} This spectrogram applies a bank of filters corresponding to tonal spacing, where each filter is equivalent to a subdivision of an octave, with central frequencies given by: 

\begin{equation}
\label{eq:cqt-0}
f_{k} = (2^{\frac{1}{b}})^{k}.f_{min}  ~~~  for  ~~ 1 \leq k \leq K
\end{equation} 
where $f_{k}$ denotes the frequency of $k$th spectral component, $f_{min}$ is minimum frequency set to 10 Hz, $b$ is the number of filters per octave as 24, and $K$ is frequency resolution of CQT, which is 128.
As the name suggest, the $Q$ value, is the ratio of central frequency to bandwidth, is constant computed as:
\begin{equation}
\label{eq:cqt-1}
Q = \frac{f_{k}}{\Delta f_{k}} = \frac{f_{k}}{f_{k+1}-f_{k}} = \left (2^{\frac{1}{b}} -1 \right )^{-1}
\end{equation}
Like STFT, CQT spectrogram is extracted using Fourier-based transformation, described as Equation (\ref{eq:cqt-2}) :
\begin{equation}
\label{eq:cqt-2}
\mathbf{CQT}[f_{k}, t] = \frac{1}{N(k)}.\sum_{n=0}^{N(k)-1}\mathbf{s}[n].\mathbf{w}[k, n-t].e^{-i2\pi \frac{nQ}{N(k)}}
\end{equation}
where:
\begin{equation}
\label{eq:cqt-3}
N(k) = Q.\frac{f{s}}{f_{k}}
\end{equation}

\begin{equation}
\label{eq:cqt-4}
\mathbf{w}[k,n] = \alpha + (1-\alpha).cos\frac{2{\pi}n}{N(k)-1}
\end{equation}
$f{s}$ is sample rate of digital input signal \(\mathbf{s}[n]\), \(\mathbf{w}[k,n]\) is window function with \(\alpha\) set to 0.54.
To generate STFT, log-Mel, MFCC, and CQT, we use a popular audio toolbox, namely Librosa~\cite{librosa_tool}.\\

\textbf{Gammatone (GAM):} Gammatone filters are designed to model the frequency-selective cochlea activation response of the human inner ear~\cite{patterson1986auditory}, in which filter output simulates the frequency response of the basilar membrane.
The impulse response is given by:
\begin{equation}
\label{eq:gammaton}
g(t) = t^{P-1}e^{-2lt\pi}cos(2ft\pi + \theta)    
\end{equation}
where \(t\) is time, \(P\) is the filter order, \(\theta\) is the phase of  the carrier, \(l\) is filter bandwidth, and \(f\) is central frequency.
The filter bank was then formulated as ERB scale~\cite{glasberg1990derivation} as:
\begin{equation}
\label{eq:gammaton}
ERB = 24.7(4.37.10^{-3}f + 1)
\end{equation}
To quickly generate Gamma spectrogram, we apply a toolbox developed by Ellis \emph{et al.}~\cite{ellis2009gammatone}, namely Gammatone-like spectrogram.
Firstly, audio signal is transformed into STFT spectra recently mentioned above.
Next, gammatone weighting $\mathbf{COE}[F_{gam}, F]$ is applied on STFT to obtain the Gamma spectrogram.
\begin{equation}
\label{eq:log-Mel}
\mathbf{GAM}[F_{gam},T] = \mathbf{COE}[F_{gam}, F].\mathbf{STFT}[F, T]
\end{equation}
where \(F_{gam}\) resolution of GAM spectrogram is Gammatone filter number of 128.

\begin{figure}[h]
	\centering
	\centerline{\includegraphics[width=\linewidth]{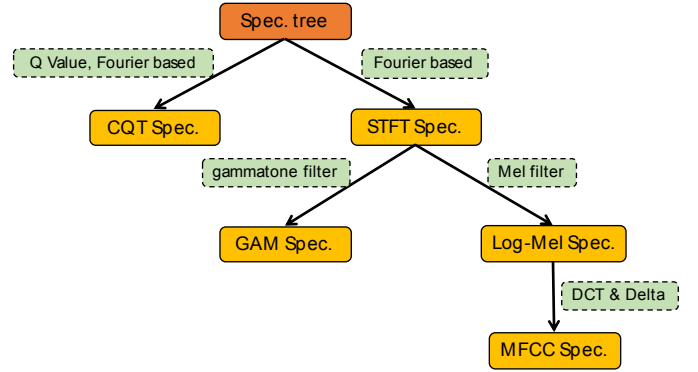}}
	\caption{Constructed spectrogram tree basing on difference of central frequencies and auditory models applied.}
	\label{fig:spectrogramTree}
\end{figure}
\begin{table}[h]
	\caption{Spectrogram combinations proposed.} 
	\vspace{0.2cm}
	\centering
	 \resizebox{0.45\textwidth}{!}{
	\begin{tabular}{l c} 
		\toprule 
		\textbf{Group of}   &  \textbf{Combinations of}  \\
		\midrule      
		Two spectrograms           & CQT+STFT,  CQT+GAM,\\     
		              & CQT+log-Mel, CQT+MFCC \\
		              \midrule    
		Three spectrograms & CQT+log-Mel+GAM, \\
		   & CQT+GAM+MFCC  \\
		    \midrule    
		Four spectrograms  & CQT+GAM+STFT+MFCC,  \\
                                           & CQT+GAM+STFT+log-Mel  \\

			   \midrule    
		Five spectrograms   & CQT+log-Mel+GAM+STFT+MFCC   \\
		\bottomrule 
	\end{tabular}}
	\label{table:combine_spec} 
\end{table}
\begin{figure*}[th]
	\centering
	\centerline{\includegraphics[width=\linewidth]{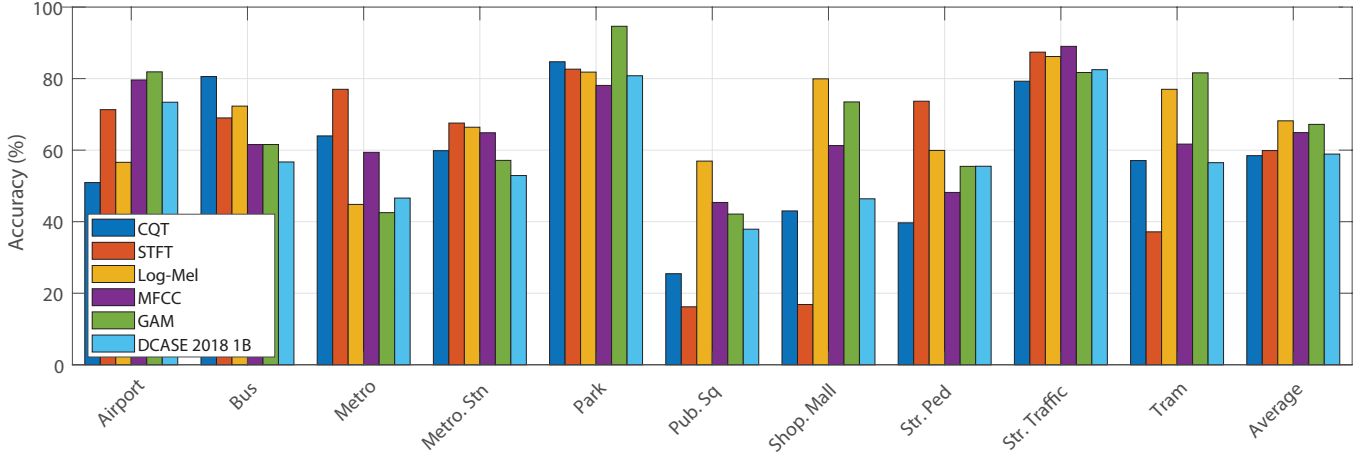}}
	\caption{Category-wise performance comparison among spectrograms on device A - DCASE 2018 Task 1B.}
	\label{fig:5spec_A}
\end{figure*}
As spectrogram formulas described, we construct a spectrogram tree as shown in Figure \ref{fig:spectrogramTree}.
Firstly, although both of CQT and STFT spectrograms are built on Fourier Transform theory, they extract different central frequencies. 
From the root tree, we therefore separate into two main branches of CQT and STFT. 
From the branch of STFT spectrogram, we continuously divide into log-Mel and GAM spectrograms due to applying different Mel and Gammatone filters, respectively.
Eventually, MFCC is an extended branch from log-Mel due to extracting DCT and Delta from this spectrogram.
It can be seen that five spectrograms proposed either extract different central frequencies or apply different auditory models.
Therefore, each spectrogram may contain its own distinct and complimentary information. 
This inspires us to conduct experiments to indicate how individual spectrograms and their combinations affect to an ASC system's performance.
Based on the tree shown in Figure \ref{fig:spectrogramTree}, we propose a variety of combinations as denoted in Table \ref{table:combine_spec}.
In particular, two-spectrogram combinations are inspired from two main branches from the root tree, each extracts specific central frequencies.
Thus, we create groups of CQT+STFT, CQT+GAM, CQT+log-Mel, and CQT+MFCC. 
There are two third-spectrogram groups of CQT+GAM+log-Mel and CQT+GAM+MFCC evaluated, which inspires from exploring different central frequencies between CQT \& STFT branches and different auditory models used among MFCC, log-Mel,  and GAM.
Inspire that applying auditory models on STFT may destroy discriminative features on this spectrogram and two spectrograms of MFCC, log-Mel may contain very similar features due to coming from same Mel filter banks, we propose two four-spectrogram combinations, which are CQT+GAM+STFT+MFCC and CQT+GAM+STFT+log-Mel.
Eventually, the combination of all five spectrograms is also evaluated. 

\subsection{Late fusion strategy to evaluate spectrogram combinations}
\label{ssec:late_fusion}

As the back-end classification works on smaller patches, the posterior probability of an entire spectrogram is computed by averaging of all patches' posterior probabilities.
Let us consider $\mathbf{P^{n}} = (\mathbf{p_{1}^{n}, p_{2}^{n},..., p_{C}^{n}})$,  with $C$ being the category number and the \(n^{th}\) out of \(N\) patches fed into learning model, as the probability of a test sound instance, then the mean classification probability is denoted as  \(\mathbf{\bar{p}} = (\bar{p}_{1}, \bar{p}_{2}, ..., \bar{p}_{C})\) where,

\begin{equation}
    \label{eq:mean_stratergy_patch}
    \bar{p}_{c} = \frac{1}{N}\sum_{n=1}^{N}p_{c}^{n}  ~~~  for  ~~ 1 \leq n \leq N 
\end{equation}

and the predicted label  \(\hat{y}\) for an individual spectrogram evaluated is determined using:

\begin{equation}
    \label{eq:label_determine}
    \hat{y} = arg max (\bar{p}_{1}, \bar{p}_{2}, ...,\bar{p}_{C} )
\end{equation}

To evaluate the combinations of spectrograms, we proposed a late fusion scheme, namely \textit{Mean} fusion.
In particular, we conduct experiments over individual spectrograms, thus obtain posterior probability of each spectrogram as  \(\mathbf{\bar{p_{s}}}= (\bar{p}_{s1}, \bar{p}_{s2}, ..., \bar{p}_{sC})\) where $C$ is the category number and the \(s^{th}\) out of \(S\) spectrograms evaluated. 
Next, the posterior probability after late fusion \(\mathbf{p_{f-mean}} = (p_{1}, p_{2}, ..., p_{C}) \) is obtained from by:

\begin{equation}
    \label{eq:mix_up_x1}
     p_{c} = \frac{1}{S} \sum_{s=1}^{S} \bar{p}_{sc} ~~~  for  ~~ 1 \leq s \leq S 
\end{equation}

Eventually, the predicted label  \(\hat{y}\) is determined by:

\begin{equation}
    \label{eq:label_determine}
    \hat{y} = arg max ({p}_{1}, {p}_{2}, ...,{p}_{C} )
\end{equation}

\subsection{Hyperparameter setting and dataset used to evaluate}
\label{ssec:hyper}

In this work, we adopt Tensorflow framework to build deep leaning models with learning rate of 0.0001, a batch size of 50, epoch number  of 100, and Adam method \cite{kingma2014adam} for learning rate optimization. 
As using mixup data augmentation, the labels are not one-hot format.
Therefore, we use Kullback-Leibler (KL) divergence loss \cite{kullback1951information} instead of the standard cross-entropy loss as shown in Equation below:

\begin{align}
\label{eq:kl_loss}
Loss_{KL}(\theta) = \sum_{n=1}^{N}\mathbf{y}_{n}\log(\frac{\mathbf{y}_{n}}{\mathbf{\hat{y}}_{n}})  +  \frac{\lambda}{2}||\theta||_{2}^{2},
\end{align}
where \(Loss_{KL}(\theta)\) is KL-loss function, $\theta$ describes the trainable parameters of the network trained, $\lambda$ denote the $\ell_2$-norm regularization coefficient experimentally set to 0.0001, \(N\) is the batch size,
$\mathbf{y_{c}}$ and $\mathbf{\hat{y}_{c}}$  are the ground-truth and the network recognized output, respectively.
Noting that we use only DCASE 2018 Task 1B dataset to analyse individual spectrograms and their combinations proposed.

\subsection{Performance comparison over spectrograms and their combinations}
\label{Feature Extraction Performance}
\begin{figure*}[h]
	\centering
	\centerline{\includegraphics[width=\linewidth]{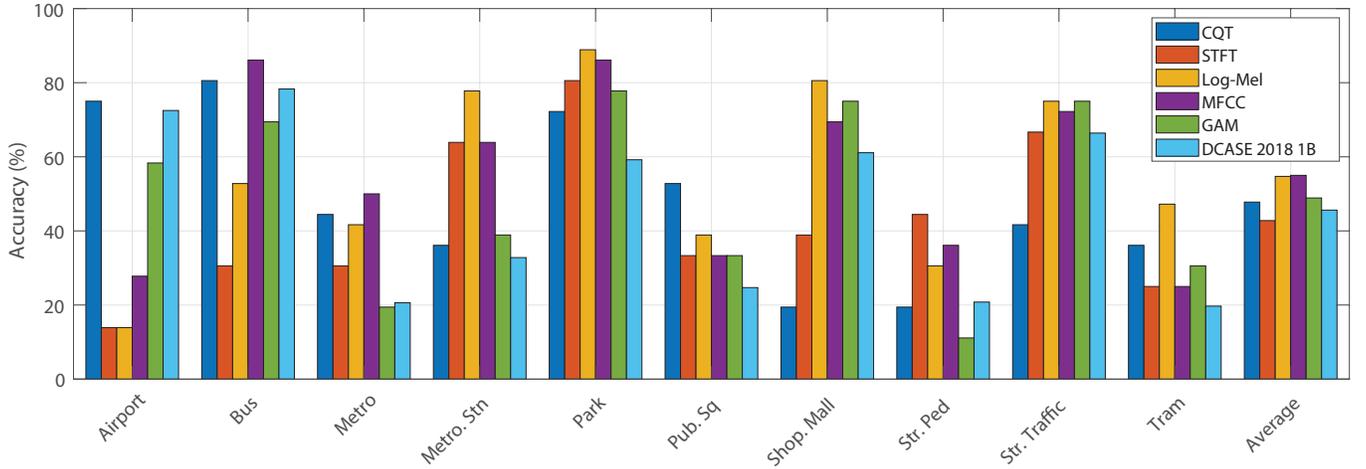}}
	\caption{ Category-wise performance comparison among spectrograms on device B\&C - DCASE 2018 Task 1B.}
	\label{fig:5spec_BC}
\end{figure*}

At initial, we equally evaluate all of five individual spectrograms by feeding them into the C-DNN network, thus show category-wise performance comparison of device A and device B\&C of DCASE 2018 Task 1B dataset.
As obtained results on device A in Figure \ref{fig:5spec_A}, log-Mel, GAM, and MFCC generally outperform STFT and CQT on most categories.
As regards the average accuracy, log-Mel and GAM stand on the top, showing competitive results of 68.2\% and 67.2\%, respectively.
Meanwhile, STFT and CQT show very low average scores compared with log-Mel and GAM, indicating a gap performance of nearly 10\%.
Noticeably, CQT spectrogram showed great performance in sound scenes belong to transportation such as \textit{Bus, Metro} and \textit{Street Traffic}.

Regarding performance on device B\&C in Figure \ref{fig:5spec_BC}, obtained results show similar with top scores of log-Mel, GAM, and MFCC. Again, CQT spectrogram still shows good accuracy rate in related-transformation categories such as \textit{Bus, Metro} and \textit{Tram}. 

In general, category-wise performance comparison of device A and device B\&C indicate that spectrograms extracted from auditory models such as GAM, log-Mel and MFCC gain high performance. 
Compare these three spectrograms to the DCASE 2018 baseline as regards Task 1B challenge (only device B\&C), they outperform the DCASE 2018 baseline over almost categories and show an improvement of 3.3\%, 9.1\%, and 9.4\% respectively in terms of the average result.

Next, we conduct experiments on two-spectrogram combinations and present obtained results in Table \ref{table:Spectrogram_combination_performance}. 
As Table  \ref{table:Spectrogram_combination_performance} shown, CQT+log-Mel achieves the greatest performance on both device A and B\&C, improving by 4\% and 6\% respectively compared to only log-Mel (the top score of individual spectrogram).
Compare CQT+log-Mel score to DCASE 2018 Task 1B baseline, it shows a significant improvement of 13.1\% and 15.2\% over device A and B\&C, respectively. 

\begin{table}[h]
	\caption{Compare individual spectrograms and their combinations to DCASE 2018 Task 1B baseline with best results (\%) in \textbf{bold}.} 
	\vspace{0.2cm}
	\centering
	\resizebox{0.5\textwidth}{!}{
		\begin{tabular}{l c c c} 
			\toprule 
			\textbf{Spectrograms}   &  \textbf{A}  &  \textbf{B\&C}  &  \textbf{A\&B\&C} \\
			\midrule      
			
			\textit{DCASE baseline}   &58.9 & 45.6 & 52.2 \\
			\midrule 
			\textit{Single spectrogram}  \\
			
			MFCC &64.9 & \textbf{55.0} &59.9\\
			STFT &59.8 & 42.7 &51.3\\
			log-Mel &\textbf{68.2} & 54.7 &\textbf{61.4}\\
			CQT &58.4 & 47.8 &53.1\\
			GAM &64.1 & 48.9 &58.1\\
				\midrule 
			\textit{Two spectrograms}  \\
			
			CQT+STFT &64.2 & 55.8 &60.0\\
			CQT+GAM  &70.9  & 53.3 &62.1 \\
			CQT+log-Mel & \textbf{72.0} & \textbf{60.8} & \textbf{66.4}\\
			CQT+MFCC  &69.8 & 58.9 &64.4 \\
			\midrule         
			\textit{Three spectrograms}  \\
			
			CQT+GAM+log-Mel &\textbf{74.1} & \textbf{62.5} &\textbf{68.3}\\
			CQT+GAM+MFCC  &71.9 & 61.1 &66.5 \\
			\midrule    
			\textit{Four spectrograms}   \\
			
			CQT+GAM+STFT+log-Mel &\textbf{74.4} & \textbf{62.5} &\textbf{68.5} \\
			CQT+GAM+STFT+MFCC   &72.7  & 60.3 &66.5 \\
			\midrule    
			\textit{All five spectrograms}   \\   
			CQT+GAM+STFT+log-Mel+MFCC   &\textbf{73.7} & \textbf{62.8}& \textbf{68.2} \\   
			
			\bottomrule 	\end{tabular}}
	\label{table:Spectrogram_combination_performance} 
\end{table}
\begin{figure*}[ht]
	\centering
	\centerline{\includegraphics[width=\linewidth]{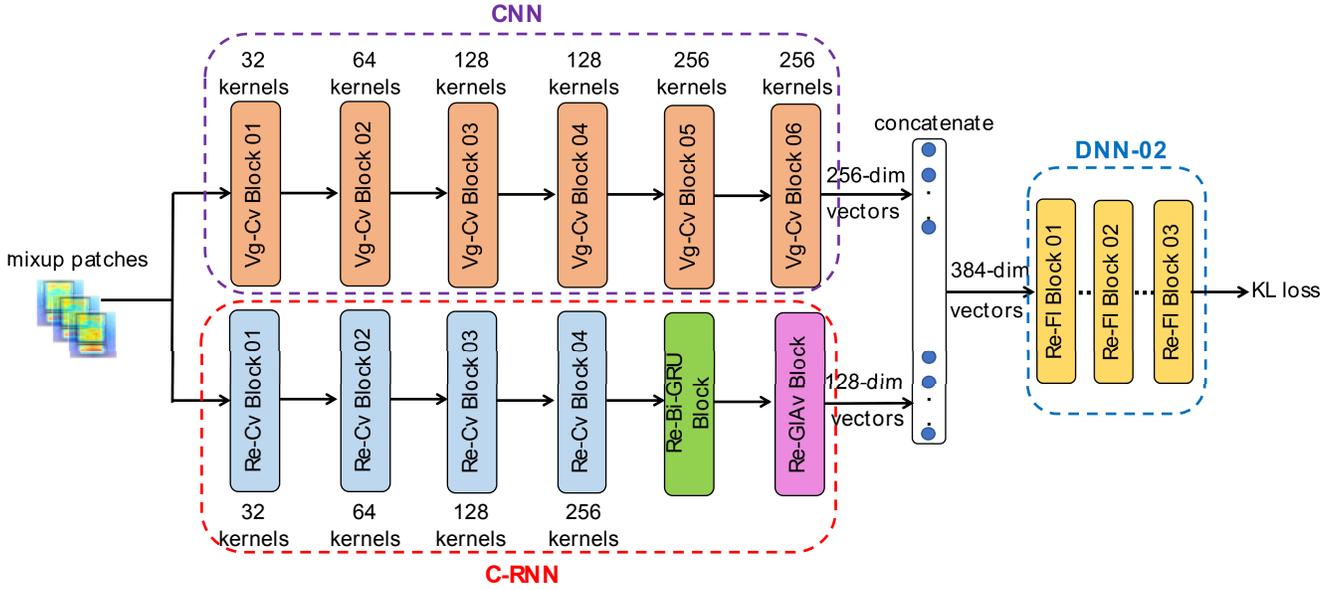}}
	\caption{Joint learning network architecture.}
	\label{fig:join_learn_model}
\end{figure*}
\begin{table*}[h]
	\caption{C-RNN network architecture.} 
	\vspace{-0.1cm}
	\centering
	\begin{tabular}{l l c} 
		\toprule 
		\textbf{Blocks}   &  \textbf{Layers}  &   \textbf{Output shape}  \\
		\midrule 
		&  Input                                                                                                   &  $128{\times}128{\times}1$ \\       
		Re-Cv Block 01 &  Bn - Cv [$4\times1$] - ReLu - Bn - Ap [$2\times1$] - Dr (10\%)  & $64{\times}128{\times}32$        \\
		Re-Cv Block 02 &  Cv [$4\times1$] - ReLu - Bn - Ap [$2\times1$] - Dr (15\%)  & $32{\times}128{\times}64$        \\
		Re-Cv Block 03 &  Cv [$4\times1$] - ReLu - Bn - Ap [$2\times1$] - Dr (20\%)  & $16{\times}128{\times}128$        \\
		Re-Cv Block 03 &  Cv [$4\times1$] - ReLu - Bn - Ap [$16\times1$] - Dr (20\%)  & $128{\times}256$        \\
		Re-Bi-GRU Block &  Bi-GRU (128 hidden states, 30\% dropout)  & $128{\times}256$        \\
		Re-GlAv Block      &  Gap & $128$        \\
		\bottomrule 
	\end{tabular} 
	\label{table:rnn_arc} 
\end{table*}
As regards three-spectrogram combinations as shown in Table \ref{table:Spectrogram_combination_performance}, two groups of CQT+GAM+MFCC and CQT+log-Mel+GAM analysed show competitive results, reporting 71.9\%, 61.1\% and 74.1\%, 62.5\% for device A and B\&C, respectively.
It indicates that log-Mel and MFCC may contain very similar features.

The results on four-spectrogram combinations of CQT+GAM+STFT+log-Mel witness  a minor increase of nearly 0.3\% and 0.2\% in terms of device A and overall respectively, compared to CQT+log-Mel+GAM, thank to the contribution of STFT. 
Meanwhile, adding STFT into CQT+GAM+MFCC only helps to improve the performance on device A a little, but makes a decrease of 0.8\% on device B\&C. As regards the result of all five spectrograms, it even has a downward trend in device A, leading to the a decrease of 0.7 \%, compared to overall accuracy in the best four-spectrogram combination of CQT+GAM+STFT+log-Mel. 

\subsection{Discussion}
\label{dicussion}

Eventually, we summary all types of spectrogram combinations, and highlight which achieve the best scores on all three devices A\&B\&C.
As results shown in Table \ref{table:Spectrogram_combination_performance}, there is a gradual increase in the accuracy rate when combination of spectrograms are applied. 
In particular, CQT+log-Mel achieves the best performance in two-spectrogram groups, with an increase of nearly 5\% compared to the best single spectrogram log-Mel. 
By adding GAM into group of  CQT+log-Mel, it helps to improve by 2\% on average.
However, a minor increase of 0.2\% is observed in the performance of four-spectrogram combination CQT+GAM+STFT+log-Mel until there is no improvement from combination of all five spectrograms. 
It can be concluded that using multiple spectrograms is effective to improve the performance, thus far exceed the DCASE 2018 Task 1B baseline.
\section{Robust deep learning framework for ASC}
\label{sec:The Proposed System}

\subsection {Join learning deep neural network architecture proposed.}

As the analysis results on spectrogram, it can be seen that using multi-spectrogram input features is effective to improve the ASC system's performance.
To further enhance, we next make effort to improve the back-end classification, thus propose a join learning model, as described in Figure \ref{fig:join_learn_model}.
As Figure \ref{fig:join_learn_model} shown, we reuse the CNN part with six Vg-Cv blocks from the C-DNN architecture.
These Convolutional blocks help to capture spatial features from spectrogram input, thus transform image patches into condensed and discriminative vectors with 256 dimension.
Additionally, we add a parallel C-RNN architecture (the lower part of Figure \ref{fig:join_learn_model}) that is used to capture structures of temporal sequences from spectrogram input.
As Table \ref{table:rnn_arc} showing the C-RNN proposed, input patches of $128\times128$ are fed into sub-blocks Cv, Bn, ReLu, Ap and Dr that are similar to those used in the CNN part.
However, we adjust settings of these sub-blocks to allow the C-RNN network be able to learn time-sequential features.
In particular, convolutional layers (Cv) with kernel size set to [$4\times1$] are applied to learn the difference between frequency banks in each temporal frame. 
Next, average pooling layers (Ap [$2/16\times1$]) is used to scale the frequency dimension of the spectrogram, but remain time resolution of 128. 
As a result, frequency dimension is scaled into 1, generating a sequence of 128-temporal frames after four Re-Cv blocks. 
Each temporal frame is represented by a 256-dimensional vector.
Next the temporal sequence is fed into bi-GRU layer in Re-Bi-GRU Block which learns the temporal sequence structure from two directions.
The output of Re-Bi-GRU Block is a matrix of $128\times256$ with 128 temporal frames and 256 dimension each frame.
Next, a Global Average Pooling layer in Re-GlAv block is applied on each temporal frame to get average results, generating a 128-dimensional vectors. 
Both output of C-RNN and CNN are thus concatenated, generate 384-dimensional vectors.
Next, these vectors are fed into a DNN-02 architecture, as shown in Table \ref{table:MLP}, configured by Fl, ReLu, Dr, and Softmax layers for classification.
Noting that output layer number $C$ depends on specific ASC task due to various datasets evaluated.
\begin{table}[h]
	\caption{ DNN-02 network architecture.} 
	\vspace{-0.1cm}
	\centering
	\resizebox{0.45\textwidth}{!}{%
		\begin{tabular}{l l c} 
			\toprule 
			\textbf{Blocks}   &  \textbf{Layers}  &   \textbf{Output shape}  \\
			\midrule 
			&  Input                                                                                                   &  $384$ \\       
			Re-Fl Block 01 &  Fl - ReLu - Dr (30\%)  & $2048$        \\
			Re-Fl Block 02 &  Fl - ReLu - Dr (30\%)  & $1024$        \\
			Re-Fl Block 03 &  Fl - Softmax & $C$        \\
			
			\bottomrule 
	\end{tabular} }
	\label{table:MLP} 
\end{table}
\subsection{Hyperparameter setting for the framework proposed.}
\label{ssec:final_setting}

The join learning model proposed is built by Tensorflow framework and reused all hyper-parameter setting from C-DNN network experiments. 
To evaluate the effect of spectrograms, we conduct experiments on the best spectrogram combinations indicated in Table \ref{table:Spectrogram_combination_performance} in Section \ref{Feature Extraction Performance}.

Additionally, we do further investigation of late fusion on accuracy. 
In particular, we compute more two fusion strategies, called \textit{Max} and \textit{Prod} fusions.
Let us consider posterior probability of each spectrogram describes as  \(\mathbf{\bar{p_{s}}}= (\bar{p}_{s1}, \bar{p}_{s2}, ..., \bar{p}_{sC})\) described in Equation (\ref{eq:mean_stratergy_patch}), where $s$ is specific spectrogram and $C$ is the number of category classified.
Next, the posterior probability of combination with \textit{Prod} strategy \(\mathbf{p_{f-prod}} = (p_{1}, p_{2}, ..., p_{C}) \) is obtained by,
\begin{equation}
\label{eq:mix_up_x1}
p_{c} = \frac{1}{S} \prod_{s=1}^{S} \bar{p}_{sc} ~~~  for  ~~ 1 \leq s \leq S 
\end{equation}
where \(S\) is the number of spectrograms combined.
The posterior probability of combination with \textit{MAX} strategy \(\mathbf{p_{f-max}} = (p_{1}, p_{2}, ..., p_{C}) \) is obtained by,
\begin{equation}
\label{eq:mix_up_x1}
p_{c} = max(\bar{p}_{1c}, \bar{p}_{2c}, ..., \bar{p}_{Sc}) 
\end{equation}
Eventually, the predicted label \(\hat{y}\) for either \textit{Max} or \textit{Prod} fusions is determined by Equation (\ref{eq:label_determine}).

As regards ASC datasets evaluated, we conduct extensive experiments on five different datasets, comprising of Litis Rouen, DCASE 2016 Task 1A, DCASE 2017 Task 1A, and DCASE 2018 Task 1A and 1B. Thus, we compare our best results to the state-of-the-art systems.

\subsection{Performance comparison on DCASE 2018 Task 1B dataset}
\label{ssec:dcase_compare}

As results obtained in Table \ref{table:re_model}, adding the C-RNN architecture into C-DNN network to create the join learning model helps to improve the performance over both device A and B\&C.
Specially, the accuracy rate increases when more spectrograms are combined.

As regards late fusion methods suggested, \textit{Prod} and \textit{Mean} are very competitive and outperform \textit{Max} fusion scheme in both C-DNN and join learning model proposed.
Noticeably, join learning models with \textit{Prod} fusion achieve the best scores for all kinds of spectrogram combinations in terms of B\&C performance.

Compare to DCASE 2018 Task 1B baseline, join learning models proposed outperform DCASE baseline on both device A and B\&C.
In particular, the best score of 67.5\% over devices B\&C, which is obtained from combination of all spectrograms CQT+log-Mel+GAM+STFT+MFCC, significantly improves the DCASE baseline by 22\%. 
It indicates that the strategy of multi-spectrogram input successfully solve the problem of mismatched devices raised in DCASE 2018 Task 1B challenge. 

\begin{table}[h!]
	\caption{\textit{Performance comparison (\%) on DCASE 2018 Task 1B dataset with the highest scores in \textbf{bold}.}} 
	\vspace{-0.1cm}
	\centering
	{
		\resizebox{0.5\textwidth}{!}{\begin{tabular}{l | c c c | c c c}
				   Architecture &     &Join Learning Model      &  &    &C-DNN    &   \\                     
				\hline          
				\textbf{Device A }   &Mean     &Prod      &Max  &Mean     &Prod      &Max    \\                     
				\hline 
				CQT+log-Mel  &72.4              &72.1               &70.9    & 72.0             &\textbf{73.0}               &70.1\\
				CQT+log-Mel+GAM   &\textbf{74.9}             &74.5                &73.4 & 74.1            &74.7               &72.3    \\
				CQT+log-Mel+GAM+STFT   &76.2            &\textbf{76.5}                &73.6  & 74.4            &74.9               &71.5\\
				CQT+log-Mel+GAM+STFT+MFCC   & 76.0            &\textbf{76.4}                &74.0  &73.7  & 74.6            &71.5    \\
				\hline 
				\hline 
				\textbf{Devices  B \& C:}    &Mean         &Prod     &Max      &Mean         &Prod     &Max  \\                     
				\hline 
				CQT+log-Mel  &62.2              &\textbf{64.7}               &60.3     &60.8              &62.2              &59.2\\
				CQT+log-Mel+GAM   &65.00             &\textbf{66.4}                &61.9     &62.5              &63.3              &59.2\\
				CQT+log-Mel+GAM+STFT   &64.44             &\textbf{66.7}                &60.6  &62.5              &60.6              &58.3\\
				CQT+log-Mel+GAM+STFT+MFCC   &65.3             &\textbf{67.5}                &64.7  &62.8              &61.9              &58.1\\
				\hline 
\end{tabular}}
	}
	\label{table:re_model} 
\end{table}

\begin{table*}[h]
	\caption{Comparison to state-of-the-art systems with best performance in \textbf{bold} (Upper part: top-ten DCASE challenges; Middle part: State-of-the-art papers; Low part: Our proposed systems using \textit{Prod} late fusion strategy ).} 
	\vspace{-0.1cm}
	\centering
	\resizebox{1\textwidth}{!}{
		\begin{tabular}{l c l c l c l c l c l c} 
			\toprule 
			\textbf{D. 2018  Task 1B}   &  \textbf{Acc.} &
			\textbf{D. 2018  Task 1A}   &  \textbf{Acc.} &
			\textbf{D. 2017 Task 1}   &  \textbf{Acc.} &
			\textbf{D. 2016 Task 1}   &  \textbf{Acc.} &
			\textbf{LITIS Roune}   &  \textbf{Acc.} & 
			
			\\(Dev. set)  & (\%) & (Dev. set)  & (\%) & (Eva. set)  & (\%) & (Eva. set)  & (\%) & (20-fold Ave.)  & (\%) \\
			\midrule      
			Baseline \cite{Mesaros2018_DCASE}        & 45.6  & 	Baseline \cite{Mesaros2018_DCASE}        & 59.7 & 	Baseline \cite{Mesaros2016_EUSIPCO}        & 74.8 & 	Baseline \cite{Mesaros2018_TASLP}        & 77.2 \\     
			
			Li \cite{li2018acoustic} & 51.7 & Li  \cite{li2018seie} & 72.9 & Zhao  \cite{zhao2017adsc} & 70.0 & Wei  \cite{dai2016acoustic} & 84.1 & Bisot  \cite{bisot2015hog} & 93.4 \\
			
			Tchorz  \cite{tchorz2018combination} & 53.9 & Jung  \cite{jung2018dnn} & 73.5 & Jung  \cite{jung2017dnn} & 70.6 & Bae  \cite{bae2016acoustic} & 84.1 & Ye  \cite{ye2015acoustic} & 96.0\\
			
			Kong  \cite{kong2018dcase} & 57.5 & Wang  \cite{wangdcase} & 73.6 & Karol  \cite{piczak2017details} & 70.6 & Kim  \cite{kim2016empirical} & 85.4 & Huy  \cite{phan2016label} & 96.4\\
			
			Wang  \cite{wang2018self} & 57.5 & Christian  \cite{roletscheck2018using} & 74.7 & Ivan  \cite{kukanov2017recurrent} & 71.7 & Takahasi  \cite{takahashi2016acoustic} & 85.6 & Yin  \cite{yin2018learning} & 96.4\\
			
			Waldekar  \cite{waldekar2018wavelet}& 57.8 & Zhang  \cite{zhang2018acoustic} & 75.3 & Park  \cite{park2017acoustic} & 72.6 & Elizalde  \cite{elizalde2016experiments} & 85.9 & Huy  \cite{phan2017improved} & 96.6\\
			
			Zhao  \cite{ren2019attention}       & 63.3 & Li  \cite{li2018acoustic} & 76.1 & Lehner  \cite{lehner2017classifying} & 73.8 & Valenti  \cite{valenti2016dcase} & 86.2 & Ye  \cite{ye2018acoustic} & 97.1\\ 
			
			Truc  \cite{nguyen2018acoustic} & 63.6& Dang  \cite{dang2018acoustic} & 76.7 & Hyder  \cite{hyder2017buet} & 74.1 & Marchi  \cite{marchi2016up} & 86.4 & Huy  \cite{phan2018beyond} & 97.8\\
			
			& & Octave  \cite{mariotti2018exploring} & 78.4 & Zhengh  \cite{weiping2017acoustic} & 77.7 & Park  \cite{park2016score} & 87.2 & Zhang  \cite{zhang2018multi} & 97.9\\
			
			& & Yang  \cite{yang2018acoustic} & 79.8 & Han  \cite{han2017convolutional} & 80.4 & Bisot  \cite{bisot2016supervised} & 87.7 & Zhang  \cite{zhang2018temporal} & 98.1\\
			
			& & Golubkov  \cite{golubkov2018acoustic} & \textbf{80.1} & Mun  \cite{mun2017generative} & \textbf{83.3} & Hamid  \cite{eghbal2016cp} & 89.7 & Huy  \cite{phan2019spatio} & 98.7\\
			
			\midrule    
			Zhao  \cite{ren2019attention}       & 63.3 & Bai  \cite{bai2019hybrid} & 66.1 & Zhao  \cite{ren2018deep} & 64.0 & Mun  \cite{mun2017deep} & 86.3 &  \\  
			
			Truc  \cite{nguyen2019acoustic} & 64.7 & Gao  \cite{gao2019adversarial} & 69.6 & Yang  \cite{yang2019kullback} & 69.3 & Li  \cite{li2017comparison} & 88.1 \\
			
			Truc  \cite{nguyen2019acousticaa} & 66.1 & Zhao  \cite{ren2019attention} & 72.6 & Waldekar  \cite{waldekar2018wavelet} & 69.9 & Hyder  \cite{hyder2017acoustic} & 88.5 \\
			
			Yang  \cite{yang2020multi} & \textbf{67.8} & Phaye  \cite{phaye2019subspectralnet} & 74.1 & Wu  \cite{wu2019enhancing} & 75.4 & Song  \cite{song2019compact} & 89.5 \\
			
			&  & Heo  \cite{heo2019acoustic} & 77.4 & Chen  \cite{chen2019audio} & 77.1 & Yin  \cite{yin2018learning} & \textbf{91.0} \\
			
			\midrule    
			log-Mel       & 58.6 & log-Mel      & 68.0 & log-Mel      & 60.3 & log-Mel       & 80.7 & log-Mel       & 97.9\\    
			
      		log-Mel+CQT       & 64.7 & log-Mel+CQT      & 70.4 & log-Mel+CQT      & 65.8 & log-Mel+CQT       & 89.2 & log-Mel+CQT       & 99.0\\  
      		   
       		log-Mel+CQT+GAM       & 66.4 & log-Mel+CQT+GAM      & 73.8 & log-Mel+CQT+GAM      & 67.3 & log-Mel+CQT+GAM       & 88.9& log-Mel+CQT+GAM       & 99.0\\    
       		 
       		log-Mel+CQT+       & 66.7 & log-Mel+CQT+      & 77.3 & log-Mel+CQT+      & 66.7 & log-Mel+CQT+       & 88.7& log-Mel+CQT+       & \textbf{99.1}\\            		
			GAM+STFT       &  & GAM+STFT      &  & GAM+STFT      &   & GAM+STFT       & &   GAM+STFT      & \\   

       		log-Mel+CQT+       & 67.5 & log-Mel+CQT+      & 77.8 & log-Mel+CQT+      & 67.0 & log-Mel+CQT+       & 88.2& log-Mel+CQT+       & \textbf{99.1}\\            		
			GAM+STFT+MFCC       &  & GAM+STFT+MFCC      &  & GAM+STFT+MFCC      &   & GAM+STFT+MFCC       & &   GAM+STFT+MFCC      & \\

			\bottomrule 
	\end{tabular}}
	\label{table:compare_state_of_the_art_system} 
\end{table*}


\subsection{Performance comparison to the-state-of-the-art systems}

We continue to evaluate our best proposed systems on various datasets, thus make a comparison to the state-of-the-art at time of writing. 
As detail presented in Table \ref{table:compare_state_of_the_art_system}, we achieve the highest accuracy of 99.1\% in LITIS Roune dataset. 
Our performance in  DCASE 2016 is 89.2\%, which lies in second position on this challenge table, and is ranked in the top-three of the state-of-the art systems. 
However, in DCASE 2017, we are out of top-ten performance in this challenge, with the figure of 67.3\%. 
As regards DCASE 2018 Task 1A dataset, the accuracy of 77.8\% obtained ranks in top four and exceed all state-of-the art systems.
Next, we again show our robustness in terms of dealing with mismatched devices issue in DCASE 2018 Task 1B.
We achieve 67.5\% in accuracy rate, outperform systems in DCASE challenge and very competitive to the top-one score in terms of state-of-the-art papers. 
It should be noted that there are inconsistencies between the reported results in the DCASE 2018 technical reports and those published in DCASE 2018 challenge website~\footnote{http://dcase.community/challenge2018/}.
The accuracy shown in Tables \ref{table:compare_state_of_the_art_system}, therefore, are collected from the original sources of technical reports.

\section{Conclusion}

This paper has presented a robust framework applying for ASC task. 
In front-end feature extraction, the idea of providing a comprehensive analysis of low-level spectrogram representation from draw audio signals able to figure out the effective types of individual spectrograms and their combinations.
As regards back-end classification, our novel join learning network based on parallel convolutional recurrent architecture has facilitated learning both spatial and temporal structural features of spectrograms. 
By approaching multi-spectrogram input and the join learning network, we achieve very competitive results compared to the state-of-the-art systems on various ASC datasets of LITIS Rouen and DCASE challenges in three consecutive years 2016, 2017 and 2018.

\vfill\pagebreak


\bibliographystyle{IEEEbib}
\bibliography{strings,refs}

\end{document}